\documentclass{PoS}
\usepackage{cite}
\usepackage{amsmath}
\usepackage{amssymb}

\title{Phase diagram of QCD with two degenerate staggered quarks}

\ShortTitle{Phase diagram of QCD}

\author{P. Cea\\
            Dipartimento di Fisica dell'Universit\`a di Bari, I-70126 Bari, Italy and INFN, Sezione di Bari, I-70126 Bari, Italy\\
            E-mail: \email{paolo.cea@ba.infn.it}}
           
\author{\speaker{L. Cosmai}\\
        INFN, Sezione di Bari, I-70126 Bari, Italy\\
        E-mail: \email{leonardo.cosmai@ba.infn.it}}
            
\author{M. D'Elia\\
            Dipartimento di Fisica dell'Universit\`a di Genova, I-16146 Genova, Italy and INFN, Sezione di Genova, I-16146 Genova, Italy\\
            E-mail: \email{massimo.delia@ge.infn.it}}

\author{A. Papa\\
            Dipartimento di Fisica dell'Universit\`a della Calabria, I-87036 Rende (Cosenza), Italy and INFN, Gruppo collegato di Cosenza, I-87036 Rende (Cosenza), Italy\\
            E-mail: \email{papa@cs.infn.it}}

\author{F.  Sanfilippo\\
           Dipartimento di Fisica, Universit\`a di Roma ``La Sapienza''  and INFN, Sezione di Roma, 
           Piazzale A. Moro 5, 00185 Roma,   Italy\\
            E-mail: \email{francesco.sanfilippo@roma1.infn.it}}

\abstract{We present preliminary results about  the critical line of QCD with two degenerate staggered quarks at nonzero temperature and chemical potential, obtained by the method of analytic continuation. As in our previous studies with different numbers of colors and flavors, we find deviations from a simple quadratic dependence on the chemical potential. 
We  comment on the shape of the critical line at real chemical potential and give an estimate of the curvature of the critical line, both for quark chemical potential and isospin chemical potential.}

\FullConference{The XXIX International Symposium on Lattice Field Theory, Lattice2011\\
		July 10-16, 2011\\
		Squaw Valley, CA, USA}

\begin{document}

\section{Introduction}
Many important issues, such as heavy ion collisions, evolution of the early universe, physics of the compact stars, are related to the knowledge of the QCD phase diagram. 
Understanding the physical phases of QCD~\cite{Philipsen:2010gj} in ordinary as well as in extreme environment of high temperature and high baryon number is therefore a major research goal.
Lattice QCD is the main nonperturbative tool to investigate the QCD phase diagram. Unfortunately lattice QCD simulations at finite quark density are plagued by the well-known sign problem.
As a matter of fact, importance sampling requires positive weights in the partition function
\begin{equation}
Z(T,\mu)= \int {\cal{D}} U e^{-S_g[U]} \det{[M(\mu)]} \,,
\end{equation}
but when $\mu\ne0$ this is no longer true, indeed $\det[M(\mu)]$ becomes complex in the case of SU(3) theory and finite quark chemical potential. There are few exceptions where $\det[M(\mu)]$ is real (positive for an even number of flavors), namely for two-color QCD, for isospin chemical potential  and for imaginary values of the quark chemical potential.
The sign problem can be addressed by means of several techniques, each one, however,  suffering from limitations.
In the last few years we approached the sign problem by the method of analytic continuation~\cite{Alford:1998sd,deForcrand:2003hx,D'Elia:2002gd}. 
To this purpose we started our investigations~\cite{Cea:2006yd,Cea:2007vt,Cea:2009ba}   from some special cases
where Monte Carlo simulations are both feasible at real and at imaginary chemical potential. The reason was to have the possibility to test the results of the analytical continuation by means of direct simulations at real chemical potential.
In the following sections we review some of our earlier results and give preliminary data for the critical line of QCD with $N_f=2$ standard staggered fermions.  

\section{Investigations about analytic continuation}
\label{Investigations}

The first lesson we learned in studying the phase diagram of two-color QCD~\cite{Cea:2007vt} was that non-linear terms in the dependence of $\beta_c$ on $\mu^2$ in general cannot be neglected. Indeed the prediction for the pseudocritical  couplings at real chemical potentials may be wrong if data  at imaginary $\mu$ are fitted according to a linear dependence in $\mu^2$. 
In Fig.~1(Left) the  discrepancy between the linear extrapolated critical line and the direct determinations of $\beta_c(\mu^2)$ 
is shown for the case of two-color QCD with $N_f=8$  degenerate staggered fermions. 
In the case of SU(3) isospin chemical potential with $N_f=8$ degenerate staggered fermions~\cite{Cea:2009ba}, instead, few interpolations can be found which correctly detect deviations from the linear behavior in $\mu^2$ at imaginary chemical potentials and
lead to consistent extrapolations to real $\mu$, in agreement with the direct determinations there.
In Fig.~1(Right) we give an example of a good fit (and the related extrapolation)  of the critical line,  using a ratio of polynomials  ${\cal{O}}(\mu^4)/{\cal{O}}(\mu^6)$. 
\begin{figure}[htbp]
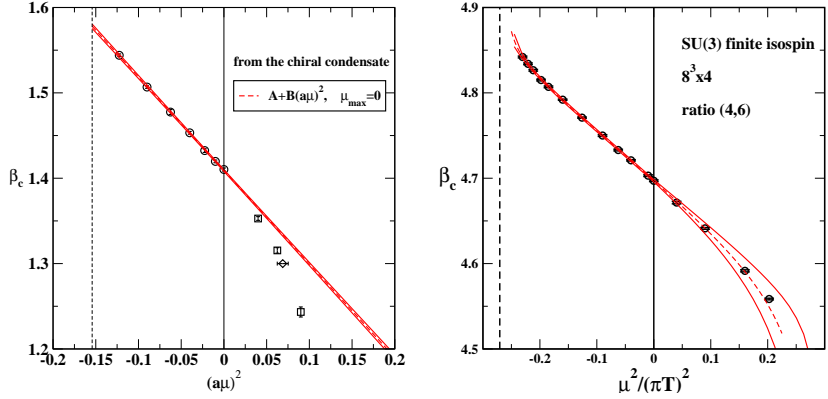
 
\centering
\begin{tabular}{cc}
\includegraphics[width=.35\textwidth,clip]{./FIGURES/SU2_Nf8_quark_linear.eps} &
\includegraphics[width=.35\textwidth,clip]{./FIGURES/SU3_Nf8_iso_ratio46.eps} 
\end{tabular} 
\caption{(Left) Critical couplings in two-color QCD at finite quark density obtained from the chiral susceptibility, together with a linear fit in $a \mu^2$ 
to data with $a \mu^2 \le 0$. 
(Right) Critical couplings in SU(3) with finite isospin density together a ratio of polynomial 
fit ${\cal{O}}(\mu^4)/{\cal{O}}(\mu^6)$.}
\label{fig1}
\end{figure}

We have also revisited~\cite{Cea:2010md} the determination of the pseudocritical line of QCD with $N_f=4$ degenerate staggered quarks at nonzero temperature and quark density by the method of analytic continuation. 
The sign problem in this case prevents us from performing simulations at real quark chemical potential. 
Notice that in this case  the linear fit is not able reproduce the data for pseudo-critical couplings even at imaginary quark chemical potentials. To derive the critical line
we exploited analytical continuation by means of polynomial or ratio of polynomials fit:
\begin{equation}
\label{fitratio}
\frac{a_0+a_1 \mu^2 + a_2 \mu^4 + a_3 \mu^6}{1+a_4 \mu^2 + a_5 \mu^4}.
\end{equation}
In Fig.~2(Left) we display the interpolation of the critical line at imaginary values of the quark chemical potential obtained by mean of a ${\cal{O}}(\mu^4)/{\cal{O}}(\mu^6)$ ratio of polynomials fit.

A nice alternative to the polynomial fits Eq.~(\ref{fitratio}) is given by what we call ``physical fit''.  
The idea is to write the interpolating function in physical units and to deduce from it the functional dependence of
$\beta_c$ on $\mu^2$, after  establishing a suitable correspondence between physical and lattice units. 
The natural, dimensionless variables of our theory are $T/T_c(0)$, where  $T_c(0)$ is the critical temperature at zero chemical potential, and $\mu/T$. The ratio $T/T_c(0)$  is deduced from the relation $T=1/(N_t a(\beta))$, where $N_t$ is the number of lattice sites in the temporal direction and $a(\beta)$   is the lattice spacing at a given $\beta$ (we discard here the dependence on the bare quark mass $am$) .  We use for $a(\beta)$ the perturbative 2-loop expression with the given number of colors and flavors. Now, adopting the 3-parameter function
\begin{equation}
\label{physical_fit}
\left[\frac{T_c(\mu)}{T_c(0)}\right]^2=\frac{1+C\mu^2/T_c^2(\mu)}
{1+A\mu^2/T_c^2(\mu)+B\mu^4/T_c^4(\mu)}\; ,
\end{equation}
we are led to the following implicit relation between $\beta_c$ and $\mu^2$
\begin{equation}
\label{a2loop}
a^2(\beta_c(\mu^2))|_{\rm 2-loop} = a^2(\beta_c(0))|_{\rm 2-loop}
\times \frac{1+A\mu^2/T_c^2 + B\mu^4/T_c^4}{1+C\mu^2/T_c^2} \,.
\end{equation}
\begin{figure}[htbp]
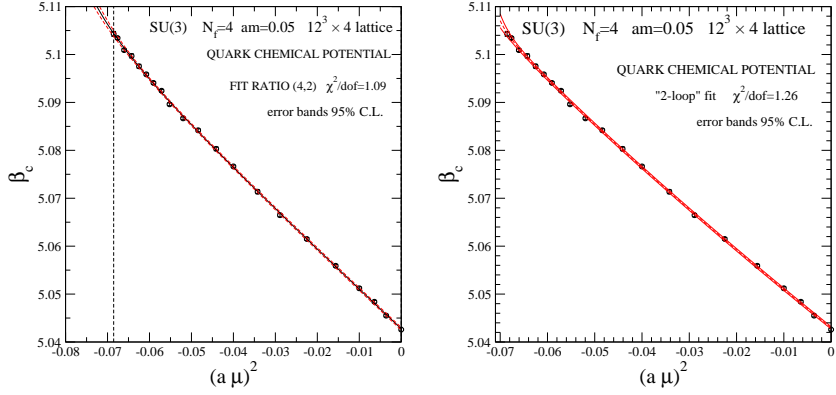
 
\centering 
\begin{tabular}{cc}
\includegraphics[width=.35\textwidth,clip]{./FIGURES/SU3_Nf4_quark_ratio42.eps} &
\includegraphics[width=.35\textwidth,clip]{./FIGURES/SU3_Nf4_quark_physical.eps} 
\end{tabular} 
\caption{SU(3), $N_f=4$, finite quark chemical potential.
Fits to the critical couplings: ratio of a 4th- to 2th-order polynomial (Left) and 2-loop ``physical fit'' according to 
Eq.~(2.3).
}
\label{fig2}
\end{figure}
In Fig.~2(Right) we display the interpolation of the critical line at imaginary values of the quark chemical potential obtains by mean of the 2-loop ``physical fit'' of Eq.~(\ref{a2loop}).

\begin{figure}[htbp] 
\centering 
\includegraphics[width=0.45\textwidth,clip]{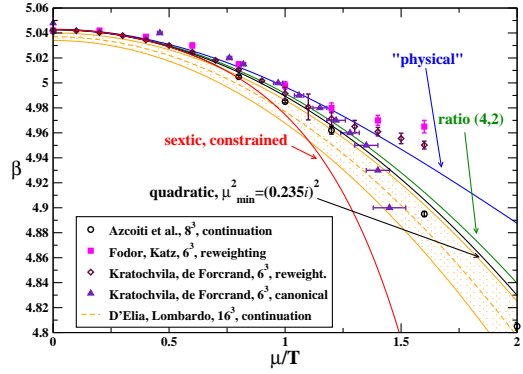} 
\caption{Comparison of our extrapolations with other determinations in the literature. For the sake of readability, our 
extrapolations have been plotted without error bands. 
Legenda: D'Elia, Lombardo, Ref.~\cite{D'Elia:2002gd}; Azcoiti et al., Ref.~\cite{Azcoiti:2005tv}; 
Fodor, Katz, Ref.~\cite{Fodor:2001au}; Kratochvila, de Forcrand, Ref.~\cite{Kratochvila:2005mk}.}
\label{fig3}
\end{figure}

Regrettably, even if we are able to have successful fits of the data at imaginary chemical potential, the
results of the extrapolation of these fits to real values of the quark chemical potential are quite consistent only up to 
 $\mu/T \sim 0.6$ (see Fig.~\ref{fig3}, where our results  are compared with data in the literature collected
 in Ref.~\cite{deForcrand:2006ec}).

\section{QCD with two degenerate staggered quarks: preliminary result}
\label{Nf2}

In this Section we present preliminary results for the critical line of QCD with two degenerate staggered flavors.
We are going to consider  the two cases of finite quark chemical potential and of finite isospin chemical potential.

We performed simulations using standard action with staggered fermion mass $am=0.05$ on a $16^3\times4$ lattice.
Simulations were done using the RHMC algorithm, properly modified for the inclusion of a finite chemical potential.  We collected typical statistics of 10 thousand trajectories.
The critical coupling at given chemical potential, $\beta_c(\mu^2)$, is determined as the value for which the susceptibility of the chiral condensate exhibits a peak. This value can be deduced by means of a Lorentzian fit to the peak or through a reweighting analysis.

\begin{figure}[htbp]
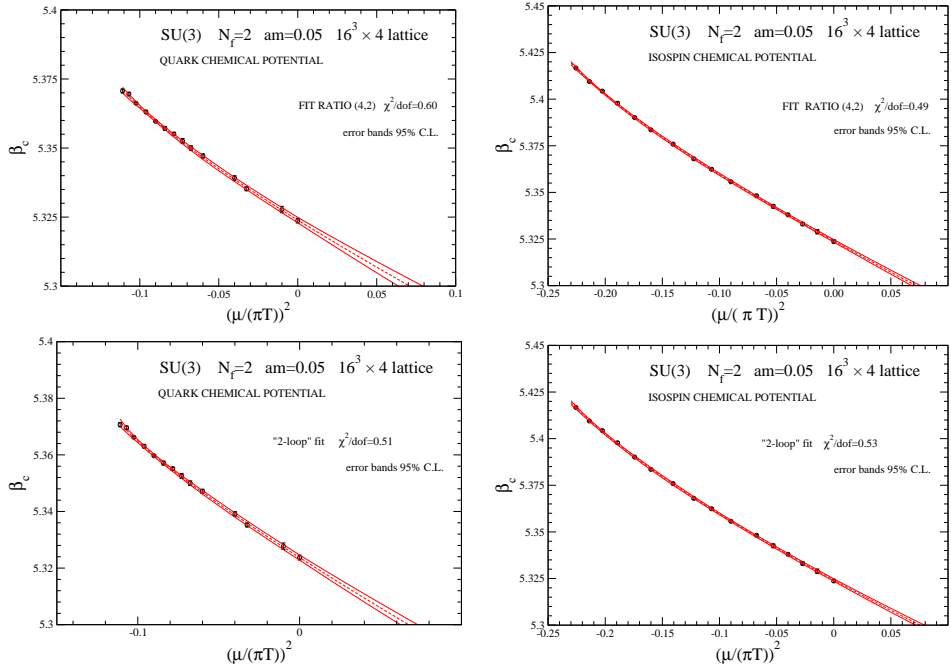
 
\centering
\begin{tabular}{cc}
\includegraphics[width=0.4\textwidth,clip]{./FIGURES/SU3_Nf2_quark_ratio42.eps} &
\includegraphics[width=0.4\textwidth,clip]{./FIGURES/SU3_Nf2_iso_ratio42.eps}  \\
\includegraphics[width=0.4\textwidth,clip]{./FIGURES/SU3_Nf2_quark_physical.eps} &
\includegraphics[width=0.4\textwidth,clip]{./FIGURES/SU3_Nf2_iso_physical.eps}  \\
\end{tabular}
\caption{(Left) SU(3) $N_f=2$, finite quark chemical potential. Fit to the critical line by means of ratio of 4th order to 2nd order polynomial (top) or by means of the ``physical fit'' Eq.~(2.3) (bottom). (Right) SU(3) $N_f=2$ isospin chemical potential. 
Ratio of polynomial fit (top) and ``physical'' fit (bottom).}
\label{fig4}
\end{figure}

We learned from our previous studies discussed in Section~2  that non-linear terms in $\mu^2$ cannot be neglected. We also learned that good fits are achieved by means of two kinds of interpolations. The first one is given by polynomial or ratio of polynomials (see Eq.~(\ref{fitratio})).
The second one consists in writing down the interpolating function in physical units and in performing a fit to the data through this implicit relation between $\beta_c$ and $\mu^2$ (see Eq.~(\ref{a2loop})).
Actually, the linear fit to the critical couplings versus imaginary values of the quark (or isospin) chemical potential
has very bad quality and we are led to exploit, even for SU(3) and $N_f=2$ staggered flavors, the functional forms
given by Eq.~(\ref{fitratio}) or by Eq.~(\ref{a2loop}).
In Fig.~\ref{fig4}  we give a sample of good interpolations obtained to data for quark chemical potential  and for isospin chemical potential   by means of the ratio of 4-th order polynomial to 2-nd order polynomial or employing 
the ``physical fit''  Eq.~(\ref{a2loop}).

The circumstance that there are significant deviations from a simple linear behavior in $\mu^2$ is evident  by looking at 
Fig.~5(Left), where a comparison between the ratio of polynomials fit and the linear fit to data at small imaginary chemical potential are displayed. The linear fit is restricted to imaginary chemical potential values such that  
$\chi^2/{\text{dof}} \lesssim 1$. This corresponds to imaginary chemical potential values with $\mu^2 \lesssim 0.1$.
\begin{figure}[htbp]
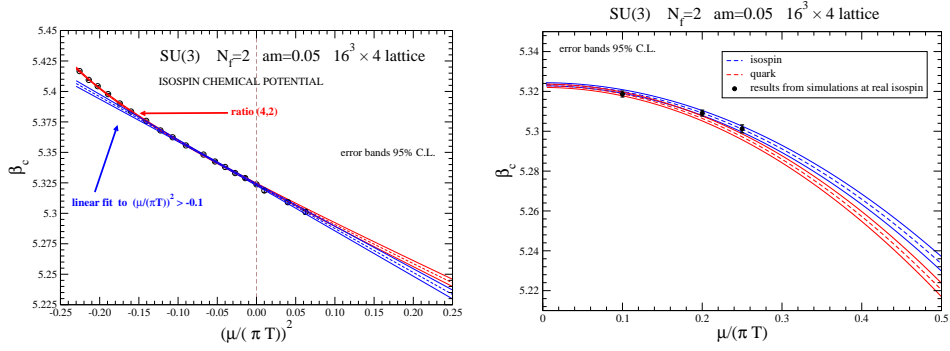
 
\centering 
\begin{tabular}{cc}
\includegraphics[width=.40\textwidth,clip]{./FIGURES/SU3_Nf2_iso_ratio42_and_linear_extrap.eps} &
\includegraphics[width=.40\textwidth,clip]{./FIGURES/SU3_Nf2_quark_iso_linear_small.eps} 
\end{tabular}
\caption{(Left) SU(3) $N_f=2$, comparison between the ratio of polynomials fit (red lines) and the linear fit to data at small imaginary chemical potential (blue lines). (Right)  Linear extrapolations starting from small values of the imaginary quark chemical potential (red lines) and small values of the imaginary isospin chemical potential (blue lines). The full dots are the results of direct numerical simulations at real values of the isospin chemical potential.}
\label{fig5}
\end{figure}
\begin{figure}[htbp]
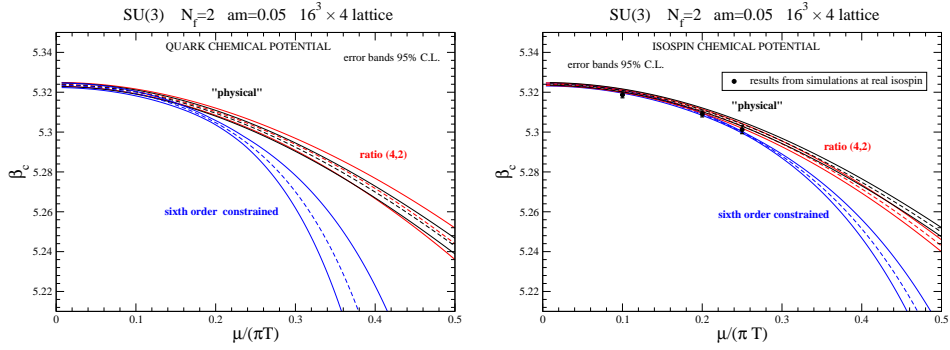
 
\centering
\begin{tabular}{cc}
\includegraphics[width=0.4\textwidth,clip]{./FIGURES/SU3_Nf2_quark_extrapolations.eps} &
\includegraphics[width=0.4\textwidth,clip]{./FIGURES/SU3_Nf2_iso_extrapolations.eps}
\end{tabular}
\caption{(Left) SU(3) $N_f=2$ isospin chemical potential. Comparison between ratio of 4th order to 2nd order polynomial fit and linear fit to small values of the imaginary isospin chemical potential. (Right) Extrapolations from linear fits at small chemical potential both for the case of isospin chemical potential (blue lines) and quark chemical potential (red lines). The full dots refer to data obtained from direct simulations at real isospin chemical potential.}
\label{fig6}
\end{figure}

In Fig.~\ref{fig6}(Left) we show a comparison between the extrapolations to real values of the quark chemical potential starting respectively from three of the more successful  interpolations of imaginary quark chemical potential data. 
If we consider the extrapolation done starting from the ``physical'' fit to the quark chemical potential data, and in the hypothesis that we can trust it down to the $T=0$ axis we can give an estimate of the critical value of $\mu$ at $T=0$. We find $\mu(T=0) = 3.284(64) T_c(0)$ to be compared with
$\mu(T=0) = 2.73(58) T_c(0)$ obtained in Ref.~\cite{PhysRevD.83.114507} with $N_f=2$ Wilson fermions.

However, while
two of the extrapolations  are quite consistent with each other, the third one (the sixth order constrained, obtained with a sixth order polynomial fit where the first term is constrained to the value obtained by a linear fit to small imaginary chemical potential) differs from the other two for values $\mu/(\pi T)$ larger than approximately $0.2$. 
In Fig.~\ref{fig6}(Right)  we compare the same extrapolations in the case of isospin chemical potential. The outcome is
almost the same.
The full circles are the values of the critical coupling from direct simulations at real isospin. The first two data are in agreement with all three different kinds of extrapolation. The third point here is in between. We are now running simulations at larger values of real isospin chemical potential.  

Finally we are able to give an estimate of the curvature of the critical line at $\mu=0$ defined as the derivative 
$d \beta_c(\mu^2)/d \mu^2$ evaluated at $\mu=0$. By means of the extrapolations from the linear fits at small chemical potential (see Fig.~\ref{fig5}(Right))
\begin{equation}
\label{linearfit}
\frac{T_c(\mu)}{T_c(0)} = 1 + a_1 \left(\frac{\mu}{\pi T}\right)^2
\end{equation}
we get the following values for the isospin and for the quark chemical potential:
\begin{eqnarray}
\label{curvatures}
a_1&=-0.470(13) \quad {\text{isospin chemical potential}}\\
a_1&=-0.522(10) \quad {\text{quark chemical potential}} \,.
\end{eqnarray}
The curvatures in the two cases differ for about 5 standard deviations.  
The curvature of the critical line  for quark chemical potential is in good agreement with $a_1=-0.500(34)$ from Ref.~\cite{deForcrand:2002ci}, where $am=0.025$. 

\section{Conclusions}
\label{Conclusions}

We have applied the method of analytic continuation to study the pseudo-critical  line $\beta_c(\mu^2)$ in QCD with $N_f=2$ degenerate staggered fermions in the case of non-zero   quark chemical potential and in the case of  non-zero isospin chemical potential.  We have estimated the curvature of the critical line. As in our previous investigations we find significant deviations from a simple  linear behavior in $\mu^2$.
Notwithstanding, there are several kinds of functions able to interpolate the critical line at imaginary values of the chemical potential, leading to extrapolations which diverge from each other at large real $\mu$.


\providecommand{\href}[2]{#2}\begingroup\raggedright\endgroup

\end{document}